\documentclass[11pt]{article}
\usepackage{arxiv}
\usepackage[utf8]{inputenc}
\usepackage[T1]{fontenc}
\usepackage{lmodern}
\usepackage{hyperref}
\usepackage{url}
\usepackage{booktabs}
\usepackage{amsfonts}
\usepackage{amsmath}
\usepackage{graphicx}
\usepackage{xcolor}
\usepackage{algorithm}
\usepackage{algorithmic}
\usepackage{tikz}
\usetikzlibrary{shapes.geometric, arrows, positioning, fit, backgrounds}
\usepackage{listings}
\usepackage{caption}
\usepackage{subcaption}
\usepackage{enumitem}
\usepackage{multirow}
\usepackage{tabularx}
\usepackage{natbib}

\hypersetup{
  colorlinks=true,
  linkcolor=blue!70!black,
  citecolor=blue!70!black,
  urlcolor=blue!70!black
}

\title{From Natural Language to PromQL: A Catalog-Driven Framework\\with Dynamic Temporal Resolution for\\Cloud-Native Observability}

\author{
  Twinkll Sisodia \\
  \texttt{twinklls@bu.edu}
}

\date{}

\begin{document}
\maketitle

\begin{abstract}
Modern cloud-native platforms expose thousands of time series metrics through systems like Prometheus, yet formulating correct queries in domain-specific languages such as PromQL remains a significant barrier for platform engineers and site reliability teams. We present a catalog-driven framework that translates natural language questions into executable PromQL queries, bridging the gap between human intent and observability data. Our approach introduces three contributions: (1)~a hybrid metrics catalog that combines a statically curated base of approximately 2{,}000 metrics with runtime discovery of hardware-specific signals across GPU vendors, (2)~a multi-stage query pipeline with intent classification, category-aware metric routing, and multi-dimensional semantic scoring, and (3)~a dynamic temporal resolution mechanism that interprets diverse natural language time expressions and maps them to appropriate PromQL duration syntax. We integrate the framework with the Model Context Protocol (MCP) to enable tool-augmented LLM interactions across multiple providers. The catalog-driven approach achieves sub-second metric discovery through pre-computed category indices, with the full pipeline completing in approximately 1.1 seconds via the catalog path. The system has been deployed on production Kubernetes clusters managing AI inference workloads, where it supports natural language querying across approximately 2{,}000 metrics spanning cluster health, GPU utilization, and model-serving performance.
\end{abstract}


\section{Introduction}
\label{sec:introduction}

Cloud-native platforms built on Kubernetes generate a substantial volume of observability signals. A moderately configured Kubernetes cluster routinely exposes upward of 5{,}000 distinct Prometheus metrics spanning cluster health, pod lifecycle, networking, storage, and, increasingly, GPU utilization for AI/ML inference workloads. The standard interface for querying these metrics is PromQL, a functional domain-specific language that, while expressive, requires practitioners to know precise metric names, label structures, histogram semantics, and time range modifiers.

In practice, a site reliability engineer investigating model-serving latency must first identify the correct metric among thousands of candidates (e.g., \texttt{vllm:e2e\_request\_latency\_seconds} versus \texttt{vllm:time\_to\_first\_token\_seconds}), understand that latency is captured in a histogram, and then construct a query using \texttt{histogram\_quantile} with appropriate rate windows and label selectors. This process is error-prone and time-consuming, especially when temporal context matters---questions like ``how has GPU temperature changed since yesterday'' or ``compare token throughput over the last 3 weeks'' require not only metric selection but also correct temporal parameterization of the resulting query.

Recent work on text-to-PromQL translation~\citep{zhang2025promcopilot} has demonstrated the feasibility of using large language models (LLMs) to bridge natural language and PromQL. PromCopilot employs a knowledge graph to capture system context and achieves 69.1\% accuracy using GPT-4 as a backbone. However, knowledge graphs require explicit construction and maintenance as the metric landscape evolves, and the approach does not address time range interpretation or hardware-specific metric discovery.

In this paper, we present a \textit{catalog-driven} framework that takes a different approach. Rather than constructing a knowledge graph, we maintain a hybrid metrics catalog: a statically curated base containing approximately 1{,}800 categorized metrics augmented by runtime discovery of hardware-specific signals. The catalog serves as the grounding layer for a multi-stage pipeline that (1)~classifies user intent into one of eight query types, (2)~routes through category-aware metric filtering, (3)~applies multi-dimensional semantic scoring, and (4)~generates syntactically correct PromQL with dynamically resolved time ranges.

Our contributions are as follows:

\begin{enumerate}[leftmargin=*, itemsep=2pt]
    \item A \textbf{hybrid metrics catalog} architecture that provides sub-second metric lookup through a static base while accommodating vendor-specific GPU metrics (NVIDIA DCGM, Intel Gaudi, AMD ROCm) via asynchronous runtime discovery, achieving catalog readiness in approximately 15\,ms.

    \item A \textbf{multi-stage NL-to-PromQL pipeline} that decomposes natural language questions through intent detection, category routing, semantic scoring, and metadata-driven query generation, with sub-second metric discovery through pre-computed category indices.

    \item A \textbf{dynamic temporal resolution} mechanism that interprets diverse natural language time expressions---from shorthand formats and relative durations to calendar references and specific dates---and maps them to appropriate PromQL duration syntax, addressing a gap in existing text-to-DSL approaches.

    \item Integration with the \textbf{Model Context Protocol} (MCP)~\citep{anthropic2024mcp} for provider-agnostic LLM tool calling, with evaluation across four LLM providers.
\end{enumerate}

\section{Background and Related Work}
\label{sec:related}

\subsection{Prometheus and PromQL}

Prometheus~\citep{prometheus2015} has become the de facto standard for metrics collection in Kubernetes environments. It scrapes time series data from instrumented endpoints and stores them with a multi-dimensional label model. PromQL provides functions for aggregation (\texttt{sum}, \texttt{avg}), rate computation (\texttt{rate}, \texttt{irate}), and distribution analysis (\texttt{histogram\_quantile}). Time range selectors (e.g., \texttt{[5m]}, \texttt{[1h]}) and offset modifiers control the temporal scope of queries. The interaction between metric types (counter, gauge, histogram, summary) and function semantics creates a combinatorial space that makes query construction non-trivial.

\subsection{Text-to-DSL and Text-to-SQL}

The problem of translating natural language to formal query languages has been extensively studied in the database domain. Spider 2.0~\citep{spider2024} and BIRD~\citep{li2024bird} benchmark text-to-SQL systems on enterprise-scale databases, revealing that even frontier LLMs achieve below 25\% accuracy on complex real-world queries. CORGI~\citep{corgi2025} extends evaluation to causal and predictive business queries, finding 33\% lower success rates compared to standard benchmarks. These results underscore the difficulty of NL-to-query translation, particularly when domain context and query semantics are complex.

In the time series domain, text-to-PromQL remains under-explored. PromCopilot~\citep{zhang2025promcopilot} is, to our knowledge, the first dedicated framework. It constructs a knowledge graph describing the cloud-native system context (services, dependencies, metrics) and combines it with LLM reasoning to generate PromQL. Using GPT-4, it achieves 69.1\% accuracy on a manually constructed benchmark of 280 questions. Our work differs in three respects: we use a catalog-driven approach instead of knowledge graphs, we handle temporal expression parsing as a first-class concern, and we support runtime metric discovery for heterogeneous hardware.

\subsection{LLMs for AIOps}

The application of LLMs to IT operations has expanded rapidly. STRATUS~\citep{stratus2025} implements a multi-agent system for autonomous site reliability engineering, coordinating specialized agents for failure detection, diagnosis, and mitigation. KubeIntellect~\citep{kubeintellect2025} provides LLM-orchestrated Kubernetes management with a 93\% tool synthesis success rate. SynergyRCA~\citep{synergyrca2025} leverages graph-based retrieval-augmented generation for root cause analysis in Kubernetes, achieving approximately 0.90 precision. An OpenTelemetry-integrated framework~\citep{otel_llm2025} reports 84.2\% reduction in mean time to resolution through multimodal analysis of metrics, logs, and traces.

These systems focus on downstream tasks (incident response, root cause analysis) and assume that the correct observability queries are available. Our work addresses the upstream challenge of formulating those queries from natural language.

\subsection{Model Context Protocol}

The Model Context Protocol (MCP)~\citep{anthropic2024mcp} standardizes tool integration for LLM-based systems. It defines a client--server architecture using JSON-RPC 2.0 where AI applications access tools, resources, and prompts through a unified interface. By February 2026, MCP had reached 97 million monthly SDK downloads and was adopted by all major LLM providers. Recent work on SMCP~\citep{smcp2026} addresses security challenges including tool poisoning and unauthorized access. We use MCP as the integration layer that allows our framework to serve both direct API consumers and LLM-based agents.

\section{System Architecture}
\label{sec:architecture}

\begin{figure}[ht]
\centering
\begin{tikzpicture}[
    node distance=0.6cm and 1.2cm,
    block/.style={rectangle, draw, rounded corners, fill=blue!8, minimum height=0.9cm, minimum width=2.8cm, align=center, font=\small},
    smallblock/.style={rectangle, draw, rounded corners, fill=green!8, minimum height=0.7cm, minimum width=2.2cm, align=center, font=\footnotesize},
    datablock/.style={rectangle, draw, rounded corners, fill=orange!10, minimum height=0.7cm, minimum width=2.2cm, align=center, font=\footnotesize},
    arrow/.style={->, >=stealth, thick},
    dashedarrow/.style={->, >=stealth, dashed, thick, gray}
]

\node[block, fill=yellow!15] (user) {User Question\\(Natural Language)};

\node[block, below=0.8cm of user] (intent) {Intent Detector\\(8 types)};
\node[block, below=0.5cm of intent] (temporal) {Temporal\\Resolver};
\node[block, below=0.5cm of temporal] (category) {Category\\Router};
\node[block, below=0.5cm of category] (scoring) {Semantic\\Scorer};
\node[block, below=0.5cm of scoring] (generator) {PromQL\\Generator};

\node[datablock, right=1.8cm of category] (static) {Static Catalog\\($\sim$1{,}800 metrics)};
\node[datablock, below=0.3cm of static] (gpu) {GPU Discovery\\(runtime)};
\node[datablock, above=0.3cm of static] (validator) {Catalog\\Validator};

\node[block, below=0.8cm of generator, fill=green!15] (output) {PromQL Query\\+ Explanation};

\node[block, right=1.8cm of scoring, fill=purple!10] (mcp) {MCP Server\\(12 tools)};
\node[smallblock, right=1.0cm of mcp] (llm) {LLM\\Providers};

\node[datablock, below=0.4cm of gpu] (prom) {Prometheus\\/ Thanos};

\draw[arrow] (user) -- (intent);
\draw[arrow] (intent) -- (temporal);
\draw[arrow] (temporal) -- (category);
\draw[arrow] (category) -- (scoring);
\draw[arrow] (scoring) -- (generator);
\draw[arrow] (generator) -- (output);

\draw[arrow] (category) -- (static);
\draw[arrow] (static) -- (scoring);
\draw[dashedarrow] (gpu) -- (static);
\draw[dashedarrow] (validator) -- (static);
\draw[arrow] (prom) -- (gpu);
\draw[arrow] (prom) -- (validator);

\draw[arrow] (mcp) -- (llm);
\draw[dashedarrow] (mcp) -- (scoring);

\begin{scope}[on background layer]
    \node[draw=gray, dashed, rounded corners, fit=(intent)(temporal)(category)(scoring)(generator), inner sep=8pt, label={[font=\footnotesize]above left:Query Pipeline}] {};
    \node[draw=gray, dashed, rounded corners, fit=(static)(gpu)(validator), inner sep=6pt, label={[font=\footnotesize]above:Hybrid Catalog}] {};
\end{scope}

\end{tikzpicture}
\caption{System architecture showing the query pipeline, hybrid metrics catalog, and MCP integration. Dashed arrows indicate asynchronous operations (GPU discovery and catalog validation occur at startup).}
\label{fig:architecture}
\end{figure}
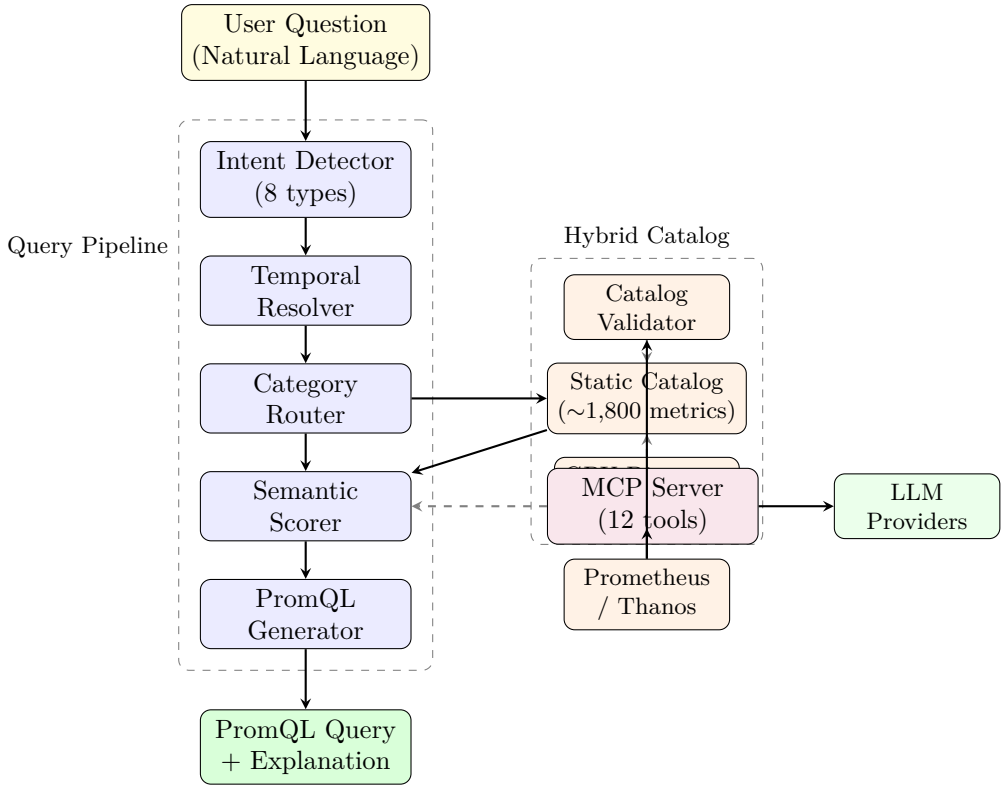

Figure~\ref{fig:architecture} illustrates the high-level architecture. The system consists of four primary components: the \textit{Metrics Catalog}, the \textit{Query Pipeline}, the \textit{Temporal Resolver}, and the \textit{MCP Tool Server}. These components operate within a FastAPI-based service deployed on Kubernetes.

\subsection{Deployment Context}

The framework is designed for Kubernetes environments running AI inference workloads. A typical deployment connects to Prometheus or Thanos for metric storage, Tempo for distributed traces, and Loki for log aggregation. The system operates either as a Kubernetes console plugin or as a standalone service, with an MCP server exposing 12 tools for metric search, query execution, and analysis.

\section{Hybrid Metrics Catalog}
\label{sec:catalog}

A central design decision is to use a \textit{pre-curated metrics catalog} as the primary knowledge source rather than a dynamically constructed knowledge graph. This section describes the catalog architecture, including its static base, runtime augmentation, and validation mechanisms.

\subsection{Static Base Catalog}

The base catalog is a JSON structure containing approximately 1{,}800 metrics organized into 17 domain categories: \textit{cluster\_health}, \textit{node\_hardware}, \textit{pod\_container}, \textit{api\_server}, \textit{etcd}, \textit{networking}, \textit{storage}, \textit{observability}, \textit{gpu\_ai}, \textit{kubelet}, \textit{scheduler}, \textit{security}, and others. Each metric entry includes:

\begin{itemize}[leftmargin=*, itemsep=1pt]
    \item \textbf{Name}: the Prometheus metric name (e.g., \texttt{container\_cpu\_usage\_seconds\_total})
    \item \textbf{Type}: counter, gauge, histogram, or summary
    \item \textbf{Help text}: the metric's description from Prometheus metadata
    \item \textbf{Priority}: High or Medium, determined by operational importance
    \item \textbf{Keywords}: up to 12 search terms generated through a five-tier process
\end{itemize}

A flat lookup table maps each metric name to its category and priority for $O(1)$ access. The catalog is bundled within the container image (approximately 840\,KB) and loads in ${\sim}15$\,ms at startup, making the service immediately ready without external dependencies.

\subsubsection{Keyword Generation}

Keywords are generated through a five-tier priority system, with each metric receiving up to 12 keywords:

\begin{enumerate}[leftmargin=*, itemsep=1pt]
    \item \textbf{Curated keywords} for well-known metrics (e.g., \texttt{DCGM\_FI\_DEV\_GPU\_UTIL} $\rightarrow$ \{``gpu utilization'', ``nvidia utilization''\})
    \item \textbf{Type-based keywords}: counters receive \{``total'', ``count'', ``rate''\}; histograms receive \{``distribution'', ``percentile'', ``p95''\}
    \item \textbf{Pattern-based expansions} from ${\sim}30$ regex rules (e.g., metrics containing ``latency'' receive \{``duration'', ``slow'', ``delay'', ``response time''\})
    \item \textbf{Name-based extraction}: splitting on underscores and colons, filtering short tokens
    \item \textbf{Help text extraction}: salient words from the Prometheus \texttt{HELP} string, filtered for stopwords
\end{enumerate}

Lower-priority tiers are truncated when the 12-keyword limit is reached, preserving curated and type-based keywords.

\subsection{Runtime GPU Discovery}

GPU metrics vary significantly across vendors and deployments. A cluster may expose NVIDIA DCGM metrics, Intel Gaudi (Habana) metrics, AMD ROCm metrics, or none at all. Static bundling would either miss vendor-specific metrics or include irrelevant entries.

We address this through asynchronous runtime discovery. At startup, a background thread queries Prometheus for all metric names, filters against vendor-specific prefix patterns (Table~\ref{tab:gpu_vendors}), and identifies the primary vendor by match count. Discovered metrics are assigned priorities using 89 vendor-specific patterns and receive keywords through a four-source generation process.

\begin{table}[ht]
\centering
\caption{Supported GPU vendors and their metric prefix patterns. Custom prefixes can be added via environment variables and are additive to defaults.}
\label{tab:gpu_vendors}
\small
\begin{tabular}{llp{4.5cm}}
\toprule
\textbf{Vendor} & \textbf{Default Prefixes} & \textbf{Key High-Priority Metrics} \\
\midrule
NVIDIA & \texttt{DCGM\_*}, \texttt{nvidia\_gpu\_*} & Utilization, temperature, power, VRAM \\
Intel & \texttt{habanalabs\_*}, \texttt{xpu\_*} & HPU utilization, energy, memory, PCIe \\
AMD & \texttt{amdgpu\_*}, \texttt{rocm\_*} & GPU busy \%, VRAM, temperature \\
Framework & \texttt{vllm:*}, \texttt{gpu\_*} & TTFT, ITL, KV cache, throughput \\
\bottomrule
\end{tabular}
\end{table}

The discovery result is merged atomically into the \texttt{gpu\_ai} category under a thread lock. If discovery fails or exceeds its 10-second timeout, the catalog continues without GPU metrics and logs a warning. This design ensures that the system is functional even in environments without GPU hardware.

\subsection{Catalog Validation}

The bundled catalog is generated from a specific Kubernetes distribution version. Different clusters may lack some catalog metrics (older versions) or expose additional metrics (newer versions). A parallel background thread validates the catalog against the live Prometheus instance at startup.

The validator builds a prefix map from existing catalog metrics, identifies stale metrics (present in catalog but absent from Prometheus) and new metrics (present in Prometheus but absent from catalog), and applies corrections atomically. New metrics are categorized using longest-prefix matching against the existing taxonomy and assigned Medium priority conservatively.

\section{Natural Language to PromQL Pipeline}
\label{sec:pipeline}

The pipeline transforms a natural language question into an executable PromQL query through four stages: intent detection, temporal resolution, category-aware metric selection, and query generation.

\subsection{Intent Detection}
\label{subsec:intent}

The first stage classifies the user's question into one of eight intent types using keyword-based pattern matching:

\begin{table}[ht]
\centering
\caption{Intent types with trigger keywords and PromQL implications.}
\label{tab:intents}
\small
\begin{tabular}{lp{4.0cm}l}
\toprule
\textbf{Intent} & \textbf{Trigger Keywords} & \textbf{PromQL Pattern} \\
\midrule
\texttt{current\_value} & ``current'', ``now'', ``what is'' & Instant query \\
\texttt{count} & ``how many'', ``total'' & \texttt{count()} \\
\texttt{average} & ``average'', ``mean'' & \texttt{avg()} \\
\texttt{percentile} & ``p95'', ``p99'', ``distribution'' & \texttt{histogram\_quantile()} \\
\texttt{top\_n} & ``top'', ``highest'', ``busiest'' & \texttt{topk()} \\
\texttt{comparison} & ``compare'', ``versus'' & \texttt{by (label)} \\
\texttt{trend} & ``over time'', ``increasing'' & \texttt{avg\_over\_time()} \\
\texttt{rate} & ``per second'', ``throughput'' & \texttt{rate()} \\
\bottomrule
\end{tabular}
\end{table}

In addition to intent, the detector extracts \textit{measurement types} (e.g., temperature, memory, latency) and domain-specific terms (e.g., TTFT, TPOT, KV cache). These features are passed downstream to influence metric scoring and query construction.

\subsection{Dynamic Temporal Resolution}
\label{subsec:temporal}

A distinctive aspect of PromQL construction is that time ranges must be explicitly specified in the query syntax. A question like ``what was the CPU usage last Tuesday'' requires not only identifying the correct metric but also resolving ``last Tuesday'' to a concrete start/end time pair and determining the appropriate PromQL rate window (e.g., \texttt{[1h]} or \texttt{[5m]}).

We introduce a multi-priority temporal resolution mechanism that maps diverse natural language time expressions to three outputs: a start timestamp, an end timestamp, and a PromQL rate syntax string. The resolver processes input through a prioritized sequence of strategies:

\begin{enumerate}[leftmargin=*, itemsep=2pt]
    \item \textbf{Shorthand formats}: Direct duration strings from UI selectors (e.g., ``15m'', ``6h'', ``7d'') are parsed via pattern matching and converted to absolute timestamps and corresponding rate syntax.

    \item \textbf{Natural language duration patterns}: Expressions of the form ``last $N$ \textit{units}'', ``$N$ \textit{units} ago'', ``in the past $N$ \textit{units}'', and ``over the last $N$ \textit{units}'' are recognized across minutes, hours, days, weeks, months, and years. The resolver maps these to both absolute time windows and appropriate PromQL rate windows that maintain query semantic correctness.

    \item \textbf{Calendar references}: Named time periods such as ``yesterday'', ``this week'', ``this month'', and specific month names (with optional year) are resolved to their calendar boundaries.

    \item \textbf{Specific date expressions}: Concrete date references are parsed using the \texttt{dateparser} library with preference for past dates, yielding day-boundary timestamps.

    \item \textbf{Explicit timestamps}: When the API caller provides start/end timestamps directly, these are used with dynamically computed rate syntax proportional to the time window.

    \item \textbf{Default fallback}: If no temporal signal is detected, a configurable default window is applied (1 hour in our deployment).
\end{enumerate}

The rate syntax selection is critical for query correctness. A 5-minute rate window (\texttt{[5m]}) applied to a 30-day query range would produce excessively sparse results, while a 30-day window applied to a 5-minute range would be meaningless. The resolver pairs each time window with an appropriate rate syntax: short windows ($< 1$h) use minute-granularity syntax, medium windows use hour-granularity, and long windows use day-granularity.

The resolved temporal information is propagated to the PromQL generator as a \texttt{time\_range\_info} structure containing the human-readable duration string, the PromQL rate syntax, and the numeric duration. This structure is then used in prompt construction to instruct the LLM on correct time range usage, and in post-processing to validate and fix generated queries.

\subsection{Category-Aware Metric Selection}
\label{subsec:selection}

Given the user's question, extracted intent, and measurement types, the system must identify the most relevant metric from the catalog. This is a retrieval problem over approximately 2{,}000 candidates.

\subsubsection{Category Hint Extraction}

A keyword-to-category mapping reduces the search space. Each of the 17 categories maintains a keyword list (e.g., \texttt{gpu\_ai} maps from \{``gpu'', ``nvidia'', ``cuda'', ``vllm'', ``ttft'', ``inference'', \ldots\}). When the user's question contains any of these keywords, the corresponding categories are activated. For questions with category hints, both High and Medium priority metrics from matching categories are returned (typically 30--80 candidates). For generic questions without hints, only High-priority metrics across all categories are returned (${\sim}350$ candidates).

\subsubsection{Multi-Dimensional Semantic Scoring}

Each candidate metric is scored along three dimensions:

\paragraph{Keyword relevance score.} A domain-specific scoring function assigns bonuses based on keyword overlap between the user's question and the metric's attributes:

\begin{equation}
S_{\text{keyword}} = \sum_{p \in \mathcal{P}} w_p \cdot \mathbf{1}[\text{match}(q, p)]
\label{eq:keyword_score}
\end{equation}

where $\mathcal{P}$ is a set of domain-specific patterns with weights $w_p$ (Table~\ref{tab:scoring_weights}), $q$ is the user query, and $\mathbf{1}[\cdot]$ is the indicator function.

\begin{table}[ht]
\centering
\caption{Semantic scoring weights for keyword pattern matching.}
\label{tab:scoring_weights}
\small
\begin{tabular}{lc}
\toprule
\textbf{Pattern Category} & \textbf{Weight} \\
\midrule
TTFT / TPOT / ITL exact match & +20 \\
GPU / CUDA / DCGM / vLLM keywords & +15 \\
Temperature keywords & +15 \\
Memory / token / cache keywords & +12 \\
CPU / network keywords & +12 \\
Latency / error keywords & +10 \\
Kubernetes patterns (pod, kube\_) & +8 \\
\bottomrule
\end{tabular}
\end{table}

\paragraph{Type relevance score.} The metric's Prometheus type is compared against the detected intent. Histogram metrics score higher for percentile intents; counters score higher for rate intents; gauges score higher for current-value intents.

\paragraph{Specificity score.} Metrics with more specific names (containing subsystem identifiers) score higher than generic metrics for targeted queries.

The total score is:
\begin{equation}
S_{\text{total}} = S_{\text{keyword}} + S_{\text{type}} + S_{\text{specificity}} + S_{\text{priority}}
\end{equation}

where $S_{\text{priority}}$ is a bonus of +15 for High-priority metrics and +5 for Medium-priority metrics. The top-scoring metric is selected for query generation.

\subsection{PromQL Generation}
\label{subsec:generation}

The final stage generates a syntactically correct PromQL query based on the selected metric, the detected intent, and the resolved time range. The generator applies intent-specific templates:

\begin{table}[ht]
\centering
\caption{PromQL generation templates by intent type and metric type. $M$ denotes the selected metric, $R$ the resolved rate syntax.}
\label{tab:promql_templates}
\small
\begin{tabular}{llll}
\toprule
\textbf{Intent} & \textbf{Counter} & \textbf{Gauge} & \textbf{Histogram} \\
\midrule
\texttt{rate} & \texttt{sum(rate(M[R]))} & \texttt{rate(M[R])} & \texttt{hq(0.95, rate(M\_b[R]))} \\
\texttt{trend} & \texttt{rate(M[R])} & \texttt{avg\_over\_time(M[R])} & \texttt{M} \\
\texttt{top\_n} & \texttt{topk(5, rate(M[R]))} & \texttt{topk(5, M)} & \texttt{topk(5, M)} \\
\texttt{comparison} & \texttt{sum by (l)(rate(M[R]))} & \texttt{avg by (l)(M)} & \texttt{hq(.95, sum by(l,le)(...))} \\
\bottomrule
\end{tabular}
\end{table}

A post-processing step validates the generated query: fixing trailing commas in label selectors, balancing parentheses, ensuring rate functions include time ranges, and converting bare range vectors into proper \texttt{rate()} calls. This syntactic repair handles common LLM generation errors and improves end-to-end query validity.

\section{MCP Integration and Multi-Provider Support}
\label{sec:mcp}

The framework is exposed through 12 MCP tools registered with a FastMCP server. These tools provide both fine-grained operations (metric search, query execution, metadata lookup) and composite operations (full smart discovery pipeline). Two consumption paths exist:

\begin{itemize}[leftmargin=*, itemsep=2pt]
    \item \textbf{LLM agents}: Call tools through an in-process adapter (\texttt{MCPServerAdapter}) that resolves tool names from the FastMCP registry and executes them directly, avoiding HTTP overhead.
    \item \textbf{Frontend/API clients}: Call tools via HTTP using JSON-RPC, allowing integration from web-based console plugins and external services.
\end{itemize}

We implement provider-specific chatbot adapters for Anthropic Claude, OpenAI GPT-4, Google Gemini, and locally deployed Llama models. Each adapter translates between the provider's native tool-calling format and the MCP tool interface. A unified \texttt{ToolExecutor} interface decouples chatbot implementations from the MCP server, following the dependency inversion principle.

The system prompt includes domain knowledge about metric types, PromQL semantics, and vLLM-specific concepts (latency decomposition, KV cache semantics, token throughput calculation). This grounding helps LLMs produce contextually appropriate explanations alongside query results.

\section{Evaluation}
\label{sec:evaluation}

We evaluate the framework through system performance measurements, catalog characteristics, illustrative query walkthroughs, and a qualitative comparison with existing approaches. The system is deployed on a Kubernetes 1.29 cluster with Prometheus/Thanos, vLLM for model serving, and NVIDIA A100 GPUs.

\subsection{Catalog Characteristics}

Table~\ref{tab:catalog_stats} summarizes the metrics catalog after startup validation against the live Prometheus instance.

\begin{table}[ht]
\centering
\caption{Metrics catalog statistics after startup validation.}
\label{tab:catalog_stats}
\small
\begin{tabular}{lr}
\toprule
\textbf{Property} & \textbf{Value} \\
\midrule
Total metrics (post-validation) & 1{,}992 \\
Domain categories & 17 \\
High-priority metrics & $\sim$350 \\
Medium-priority metrics & $\sim$1{,}650 \\
Keywords per metric (max) & 12 \\
GPU vendor prefixes monitored & 3 (NVIDIA, Intel, AMD) \\
GPU high-priority detection patterns & 89 \\
Curated keyword entries (GPU) & 59 metrics, 210 keywords \\
MCP tools registered & 12 \\
LLM providers supported & 4 \\
\bottomrule
\end{tabular}
\end{table}

The catalog covers the standard Kubernetes metric space comprehensively. After validation, stale metrics are pruned and newly discovered metrics from the live cluster are incorporated, ensuring the catalog reflects the actual deployment.

\subsection{System Performance}

Table~\ref{tab:performance} reports latency measurements for key operations, collected from production deployments.

\begin{table}[ht]
\centering
\caption{System performance measurements from production deployment.}
\label{tab:performance}
\small
\begin{tabular}{lr}
\toprule
\textbf{Operation} & \textbf{Latency} \\
\midrule
Cold start (catalog load from JSON) & $\sim$15\,ms \\
Cached catalog access (singleton) & $\sim$0.05\,ms \\
Category filtering & 3--5\,ms \\
GPU discovery (async, background) & 1--2\,s \\
Catalog validation (async, background) & 1--2\,s \\
Smart metric discovery (catalog path) & $\sim$1.1\,s \\
Smart metric discovery (API fallback) & $\sim$3.7\,s \\
Temporal resolution & $<$ 1\,ms \\
\bottomrule
\end{tabular}
\end{table}

The catalog path is 3.4$\times$ faster than the API fallback path because it avoids per-metric Prometheus metadata API calls. GPU discovery and catalog validation run asynchronously in background daemon threads and do not affect query latency after the initial startup window. The server is ready to handle queries within $\sim$15\,ms of process start; GPU-specific queries become available 1--2\,s later when discovery completes.

\subsection{Illustrative Query Walkthroughs}

We present three end-to-end examples that demonstrate the pipeline across different complexity levels and the role of temporal resolution.

\paragraph{Example 1: Simple query with domain-specific abbreviation.}

\begin{quote}
\textit{``What is the TTFT for my vLLM deployment?''}
\end{quote}

\begin{enumerate}[leftmargin=*, itemsep=1pt]
    \item \textbf{Intent}: \texttt{current\_value} (triggered by ``what is'')
    \item \textbf{Temporal}: No time expression detected; defaults to last 1 hour, rate syntax \texttt{[1h]}
    \item \textbf{Category}: Keywords ``ttft'' and ``vllm'' activate \texttt{gpu\_ai} category, reducing candidates from $\sim$2{,}000 to $\sim$50
    \item \textbf{Scoring}: \texttt{vllm:time\_to\_first\_token\_seconds} scores highest (+20 for TTFT exact match, +15 for vLLM keyword, +15 priority bonus)
    \item \textbf{Output}: \texttt{histogram\_quantile(0.95, sum(rate(vllm:time\_to\_first\_token\_seconds\_bucket[1h])) by (le))}
\end{enumerate}

\paragraph{Example 2: Temporal query with relative duration.}

\begin{quote}
\textit{``How has GPU temperature changed over the last 6 hours?''}
\end{quote}

\begin{enumerate}[leftmargin=*, itemsep=1pt]
    \item \textbf{Intent}: \texttt{trend} (triggered by ``changed'' and ``over'')
    \item \textbf{Temporal}: ``last 6 hours'' parsed to 6-hour window, rate syntax \texttt{[6h]}
    \item \textbf{Category}: ``GPU'' and ``temperature'' activate \texttt{gpu\_ai}, scoring boosts temperature-related metrics (+15)
    \item \textbf{Scoring}: \texttt{DCGM\_FI\_DEV\_GPU\_TEMP} selected (gauge type matches trend intent)
    \item \textbf{Output}: \texttt{avg\_over\_time(DCGM\_FI\_DEV\_GPU\_TEMP[6h])}
\end{enumerate}

\paragraph{Example 3: Complex query with calendar reference.}

\begin{quote}
\textit{``Compare token throughput across models since yesterday''}
\end{quote}

\begin{enumerate}[leftmargin=*, itemsep=1pt]
    \item \textbf{Intent}: \texttt{comparison} (triggered by ``compare'')
    \item \textbf{Temporal}: ``yesterday'' resolved to 24-hour window ending now, rate syntax \texttt{[1d]}
    \item \textbf{Category}: ``token'' and ``throughput'' activate \texttt{gpu\_ai}
    \item \textbf{Scoring}: \texttt{vllm:generation\_tokens\_total} selected (counter type, +12 for token keyword)
    \item \textbf{Output}: \texttt{sum by (model\_name)(rate(vllm:generation\_tokens\_total[1d]))}
\end{enumerate}

These examples illustrate how temporal resolution and category routing work together: the resolved rate syntax (\texttt{[1h]}, \texttt{[6h]}, \texttt{[1d]}) is propagated into the generated PromQL, ensuring that the time window in the query matches the user's intent.

\subsection{Comparison with Existing Approaches}

Table~\ref{tab:comparison} provides a qualitative comparison with PromCopilot~\citep{zhang2025promcopilot}, the closest related system.

\begin{table}[ht]
\centering
\caption{Qualitative comparison with PromCopilot. PromCopilot reports 69.1\% semantic accuracy using GPT-4 on their 280-question benchmark.}
\label{tab:comparison}
\small
\begin{tabular}{p{3.2cm}p{3.8cm}p{3.8cm}}
\toprule
\textbf{Aspect} & \textbf{PromCopilot} & \textbf{Our Framework} \\
\midrule
Knowledge source & Knowledge graph (manually constructed) & Hybrid catalog (static + runtime discovery) \\
Setup overhead & Requires KG construction per system & Zero-config; catalog bundled, validated at startup \\
Time range handling & Not addressed & Multi-priority NL temporal resolution \\
Hardware support & General cloud metrics & Multi-vendor GPU discovery (NVIDIA, Intel, AMD) + vLLM inference metrics \\
LLM providers & GPT-4 only & 4 providers via MCP \\
Query generation & LLM + KG reasoning & Intent-driven templates with post-processing repair \\
Metric count & Not specified & $\sim$2{,}000 categorized with keywords \\
Startup latency & Not reported & $\sim$15\,ms (catalog), 1--2\,s (GPU discovery, async) \\
\bottomrule
\end{tabular}
\end{table}

The primary trade-off is expressiveness versus maintainability. The knowledge-graph approach can represent service dependencies and causal relationships, enabling queries that require cross-service reasoning. The catalog approach cannot represent these relationships but requires no construction step, updates automatically, and provides faster metric lookup through pre-computed category indices and keyword maps.

\section{Discussion}
\label{sec:discussion}

\paragraph{Catalog versus knowledge graph.} The catalog approach trades expressiveness for maintainability. A knowledge graph can represent service dependencies and causal relationships, which our catalog cannot. However, the catalog requires no separate construction step, updates automatically through validation, and scales trivially to new metric sources. In our deployment, the catalog's category-based routing proved sufficient for the vast majority of single-metric queries. Multi-metric queries that require understanding service dependencies (e.g., ``which service is causing latency in the payment pipeline'') remain better served by knowledge-graph approaches.

\paragraph{Temporal resolution limitations.} The resolver handles the most common temporal patterns in observability contexts but struggles with relative references that require world knowledge (``since the last deployment'', ``during the outage''). These expressions require integration with event systems (deployment logs, incident records) that is beyond the current scope.

\paragraph{Failure modes.} Through deployment experience, we observe three primary failure modes: (1)~metric ambiguity, when multiple metrics could reasonably satisfy the question and the scorer selects a plausible but unintended candidate, (2)~incorrect PromQL function selection for complex intents, particularly when the metric type (histogram vs.\ counter) interacts non-obviously with the intent, and (3)~label selector errors, where the correct metric is identified but the generated query applies wrong or missing label filters. Temporal resolution errors occur primarily on ambiguous expressions that require world knowledge (e.g., ``since the last deployment'').

\section{Conclusion}
\label{sec:conclusion}

We have presented a catalog-driven framework for translating natural language questions into PromQL queries in cloud-native observability environments. The hybrid metrics catalog, combining a statically curated base of approximately 2{,}000 metrics with runtime hardware discovery across three GPU vendors, provides a practical alternative to knowledge graphs for metric grounding. The multi-stage pipeline---intent detection, temporal resolution, category-aware metric selection, and template-driven generation---achieves sub-second metric discovery and produces syntactically valid PromQL across eight distinct intent types. The dynamic temporal resolution mechanism addresses a gap in existing text-to-DSL approaches by handling diverse natural language time expressions and mapping them to appropriate query syntax.

The system is deployed on production Kubernetes clusters managing AI inference workloads, demonstrating the viability of LLM-augmented observability for production environments. Future work includes extending the temporal resolver to handle event-relative references, incorporating multi-metric query planning for complex diagnostic scenarios, and developing automated benchmark generation from production query logs.

\section*{Acknowledgments}


\bibliographystyle{plainnat}
\bibliography{references}

\end{document}